\documentclass[fleqn,twoside]{article}
\usepackage{espcrc2}
\usepackage{epsf}

\title{Low Energy Chiral Lagrangian Parameters for Scalar and
Pseudoscalar Mesons}

\author{
    W.\ Bardeen$^{a}$,
    E.~Eichten\thanks{Presenter}\address{Fermilab, PO Box 500, Batavia, IL 60510}%
and 
    H.~Thacker\address{Dept. of Physics, University of Virginia, 
Charlottesville, VA 22901}
}

\begin{document}

\begin{abstract}
We present results of a high-statistics study of scalar and pseudoscalar meson 
propagators in quenched QCD at two values of lattice spacing, $\beta = 5.7$ 
and 5.9, with clover-improved Wilson fermions. The study of the chiral limit is 
facilitated by the pole-shifting ansatz of the modified quenched approximation. 
Pseudoscalar masses and decay constants are determined as a function of quark mass 
and quenched chiral log effects are estimated. A study of the flavor singlet 
$\eta'$ hairpin diagram yields a precise determination of the $\eta'$
mass insertion. The corresponding value of the quenched chiral log 
parameter $\delta$ is compared with the observed QCL effects. Removal of QCL 
effects from the scalar propagator allows a determination of the mass of the 
lowest lying isovector scalar $q\bar{q}$ meson. 
\vspace{1pc}
\end{abstract}

\maketitle

\section{Quenched Chiral Perturbation Theory}

The lowest order chiral Lagrangian supplemented by the
chiral symmetry breaking terms $L_{5}$ of
O($p^{2}m_{\pi}^{2}$) and $L_{8}$ of
O($m_{\pi}^4$) (which model the mass-dependence of the
chiral slope for the meson decay constants) is given by\cite{chlogs}:.
\begin{eqnarray}
\label{eq:chilag}
\lefteqn{{\cal L}}~&& \hspace{-1em}=~ \frac{f^2}{4}{\rm Tr}(\partial_{\mu}U^{\dagger}\partial^{\mu}U) + 
 \frac{f^2}{4}{\rm Tr}(
\chi^{\dagger}U+U^{\dagger}\chi) \nonumber \\
   &&+~L_5 {\rm Tr}(\partial_{\mu}U^{\dagger}\partial^{\mu}U(\chi^{\dagger}U+U^{\dagger}
        \chi)) \\
   &&+~L_8 {\rm Tr}(\chi^{\dagger}U\chi^{\dagger}U+U^{\dagger}\chi U^{\dagger}\chi) 
         + {\cal L}_{\rm hairpin} \nonumber
\end{eqnarray}
where
\begin{equation}  
 {\cal L}_{\rm hairpin} \equiv -\frac{1}{2} m_{0}^{2}\frac{f^2}{8}(i{\rm Tr}\ln(U^{\dagger})-i{\rm Tr}\ln(U))^2
\end{equation} 

We use explicit formulas for the pseudoscalar masses, 
and pseudoscalar and axial vector decay constants, consistent through 
order $p^4$ (for details see \cite{chlogs}).
The coefficients $L_{5}, L_{8}$ follow the notation of 
Gasser and Leutwyler \cite{Leutwyler}. 

\section{Pseudoscalar masses and chiral parameters}

The expression for the pseudoscalar mass squared
up to first order in the hairpin mass, $L_5$ and $L_8$ is:

\begin{eqnarray}
\label{massformula}
\lefteqn{M^{2}_{ij}}&& =~\chi_{ij} (1+\delta I_{ij})\{1+  
       \frac{1}{f^2}8(2L_8 -L_5)\chi_{ij} [1 
\nonumber \\
   && ~+ \delta (\tilde{I}_{ij} + I_{ij})] 
       +\frac{1}{f^2}8 L_5 \chi_{ij} \delta \tilde{J}_{ij} \} 
\end{eqnarray}
where $r_{0}$ is a slope parameter,
$\chi_i =  2r_{0}m_{i}$,  
$\chi_{ij} = (\chi_i + \chi_j)/2$ and
$m_{i} \equiv \ln(1+ 1/(2\kappa_{i}) - 1/(2\kappa_{c}))$.
Here $\tilde{I}_{ij} = (I_{ii}\chi_i+I_{jj}\chi_j)/\chi_{ij}$ and
$\tilde{J}_{ij} = J_{ij}/\chi_{ij}$. The loop integrals $I_{ij}$
and $J_{ij}$ are defined in \cite{chlogs}.

For the pseudoscalar decay constants:
\begin{eqnarray}
\label{eq:fpi}
\lefteqn{f_{P;ij}}&&=~\sqrt{2}fr_{0}(1+ 0.25\delta(I_{ii}+I_{jj}+2I_{ij}))\{1
\nonumber \\
  &&~+~\frac{4}{f^2}(4L_8 - L_5)\chi_{ij}[1+\delta(\tilde{I}_{ij} + I_{ij})]
\nonumber \\
  &&~-~\frac{4}{f^2}L_{5}\delta \chi_{ij} [(\frac{\tilde{I}_{ij}}{2} 
          + I_{ij}) - (\tilde{J}_{ii}+\tilde{J}_{jj})]\} 
\end{eqnarray}

While for the axial vector decay constants:
\begin{eqnarray}
\label{faformula}
\lefteqn{f_{A;ij}}&&~=~ \sqrt{2}f(1+ 0.25\delta(I_{ii}+I_{jj}-2I_{ij}))
\nonumber \\
     &&~\{~ 1 + \frac{4}{f^2}\chi_{ij} L_{5}[1 + \delta(2I_{ij} - \tilde{I}_{ij}) \} 
\end{eqnarray}

\begin{figure}
\epsfxsize = 0.45\textwidth
\centerline{\epsfbox{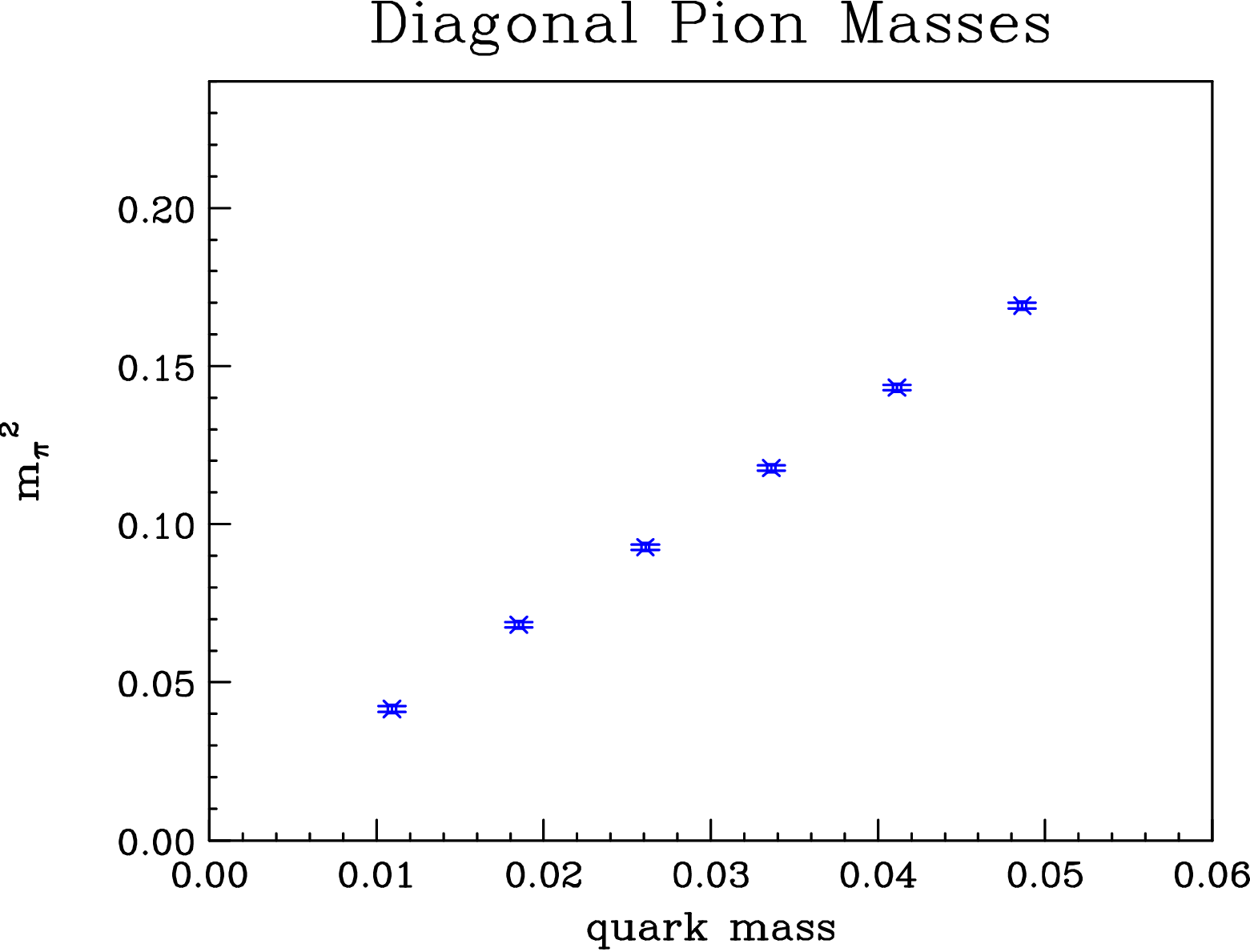}}
\end{figure}
\begin{figure}
\epsfxsize = 0.45\textwidth
\centerline{\epsfbox{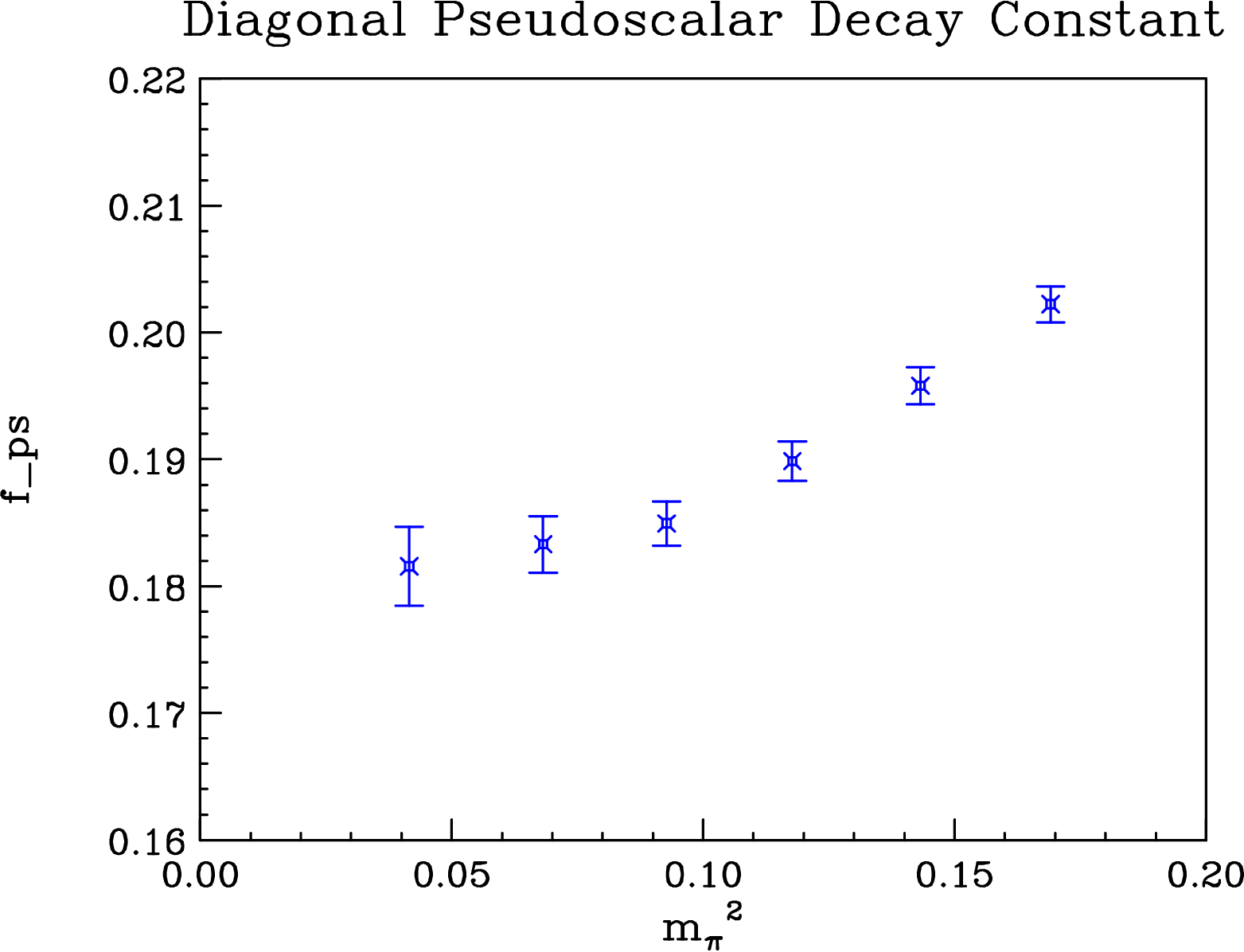}}
\end{figure}
\begin{figure}
\epsfxsize = 0.45\textwidth
\centerline{\epsfbox{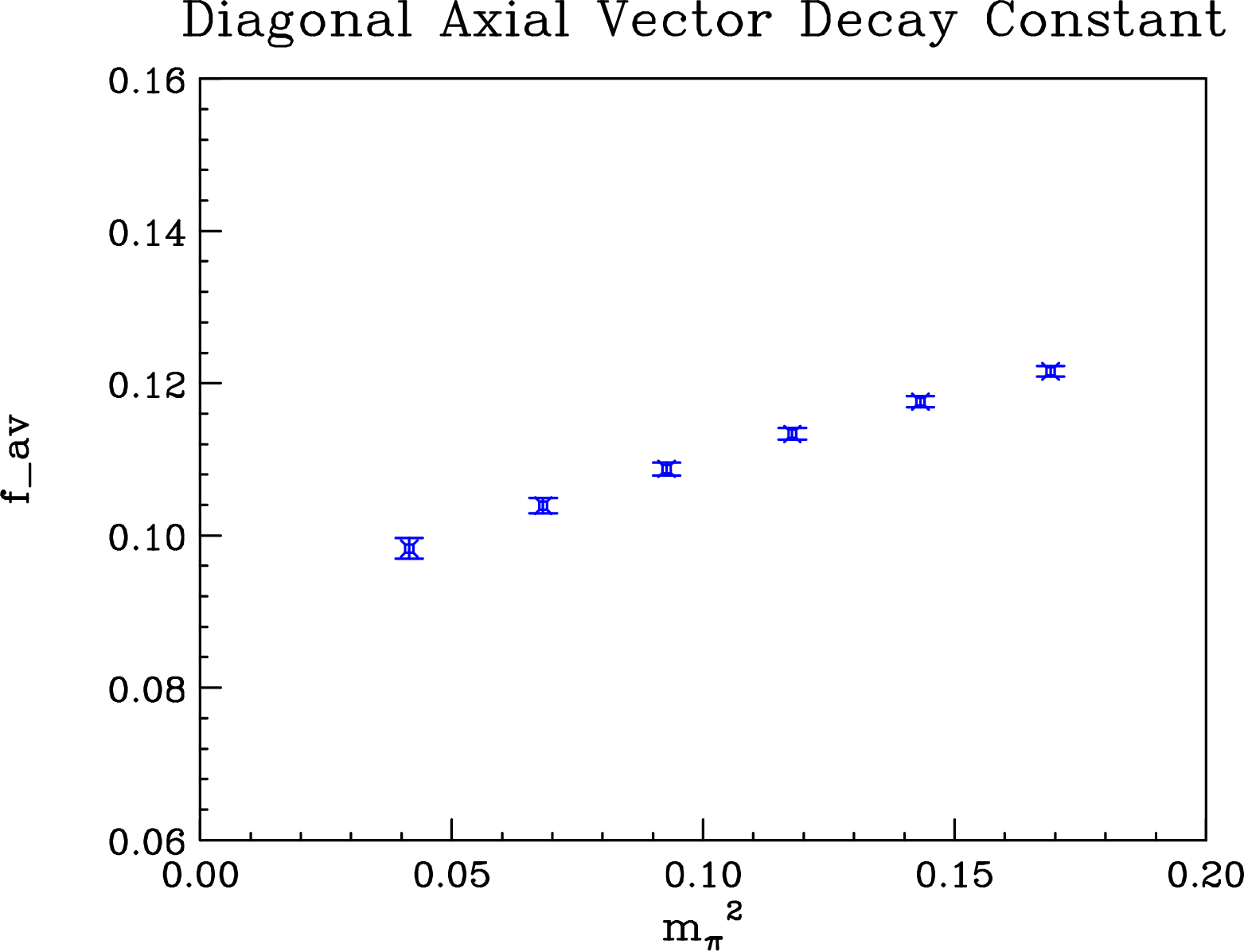}}
\caption{$m_{\pi}^2$, $f_{\rm ps}$ and $f_{\rm ax}$ versus diagonal masses.}
\label{fig:diag}
\end{figure}
\begin{figure}
\epsfxsize = 0.45\textwidth
\centerline{\epsfbox{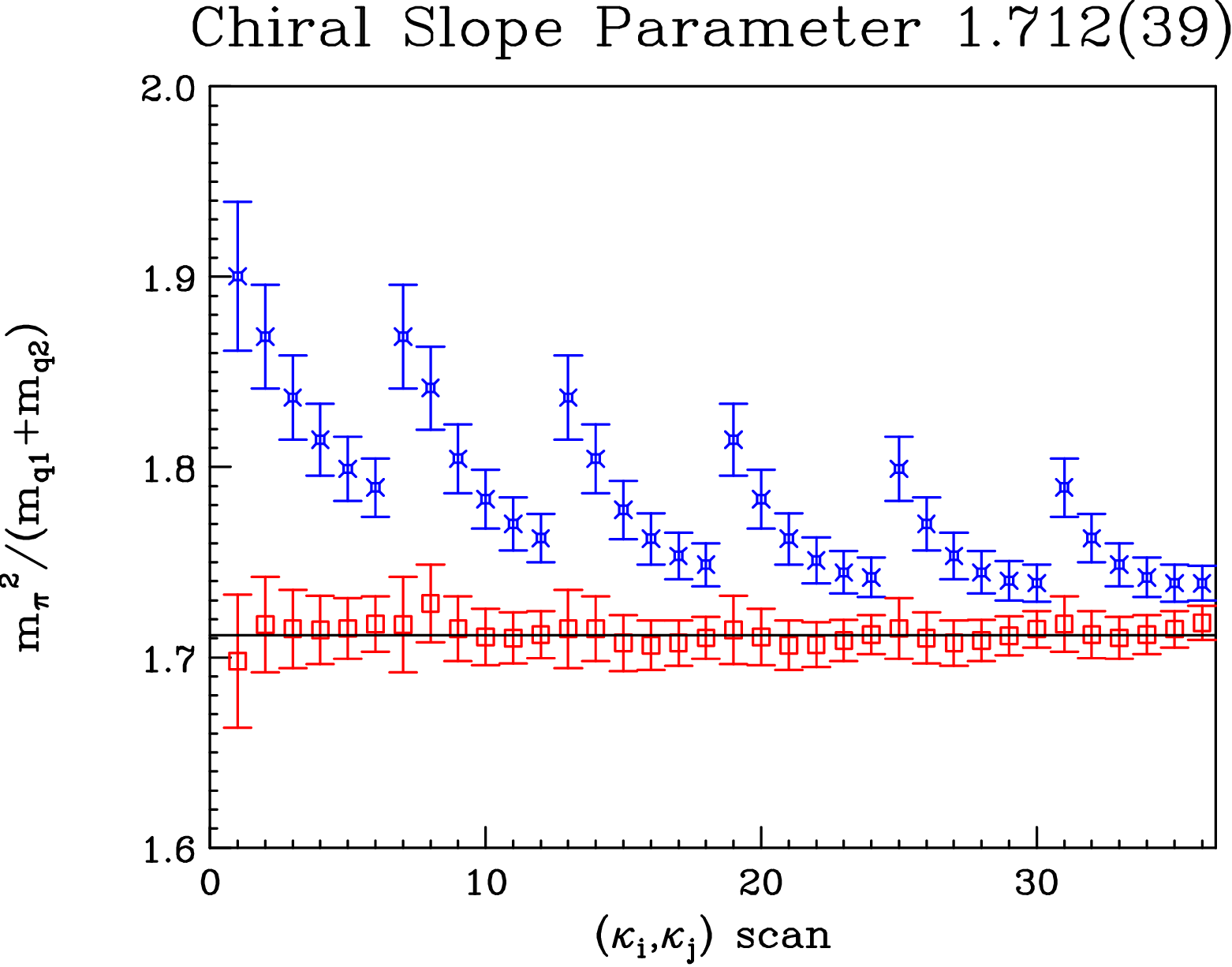}}
\end{figure}
\begin{figure}
\epsfxsize = 0.45\textwidth
\centerline{\epsfbox{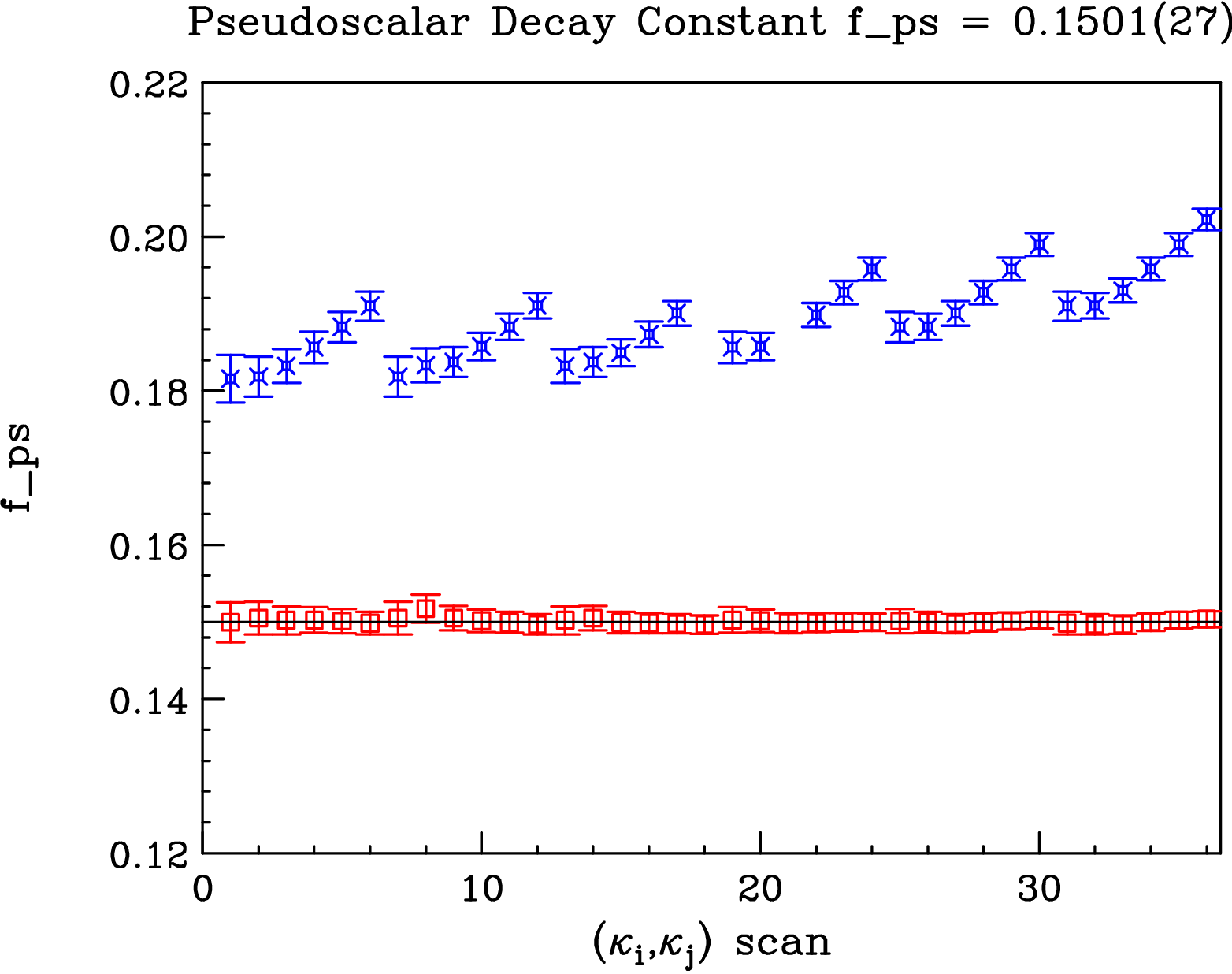}}
\end{figure}
\begin{figure}
\epsfxsize = 0.45\textwidth
\centerline{\epsfbox{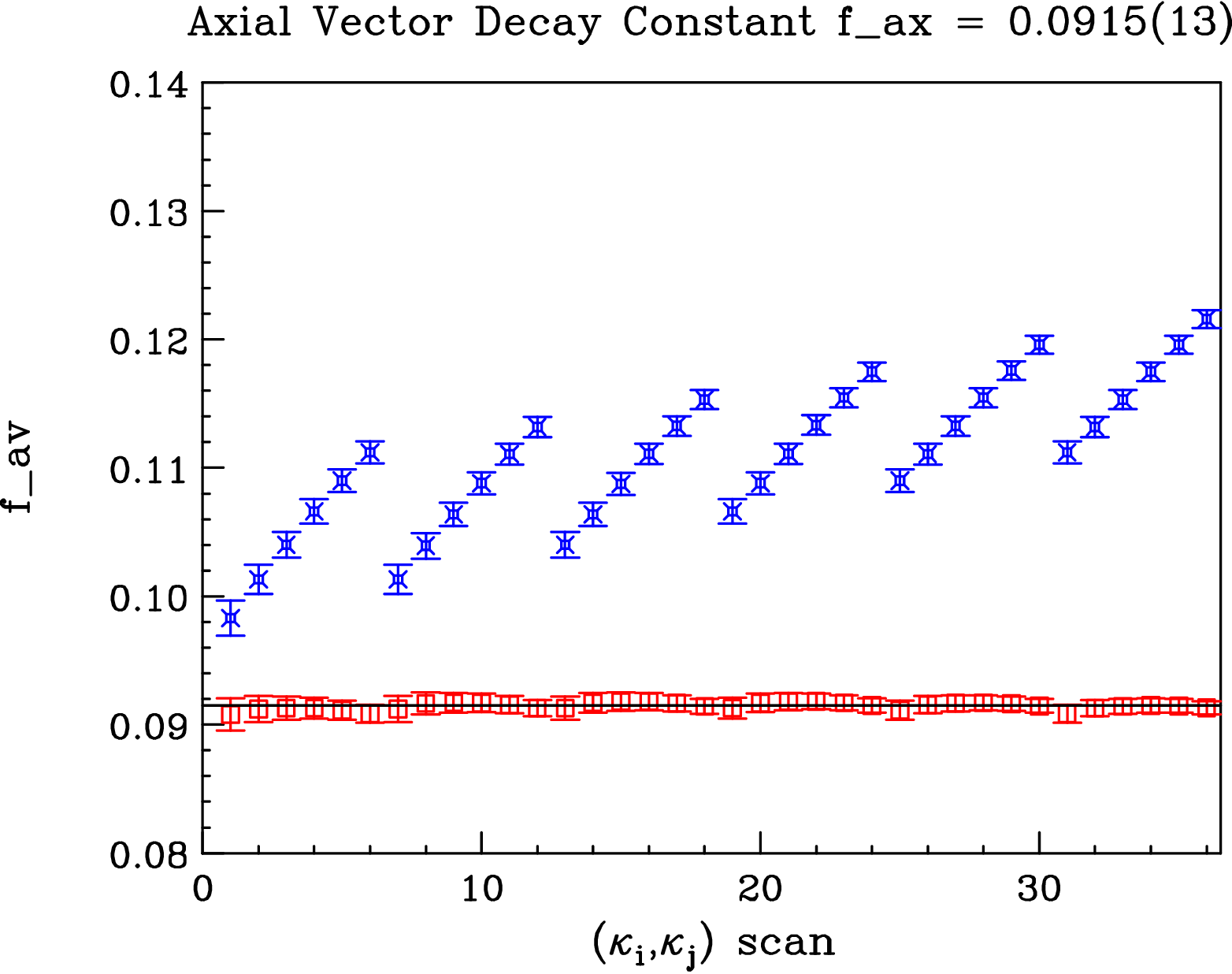}}
\caption{$m_{\pi}^2$, $f_{\rm ps}$ and $f_{\rm ax}$ and ratios to chiral fits.}
\label{fig:ratio}
\end{figure}

We study 350 configurations on a $16^3\times 32$ lattice with
$\beta =5.9$. Clover fermions ($C_{SW} = 1.50$) and 
the MQA procedure\cite{MQA} was used. 
We consider six $\kappa$ values ($.1397, .1394, .1391, .1388, .1385, .1382$) 
with $\kappa_c = .140143$. The fits are shown in Figures \ref{fig:diag}-\ref{fig:ratio}.

\section{Hairpins and Scalar Propagators}

\begin{figure}
\epsfxsize = 0.42\textwidth
\centerline{\epsfbox{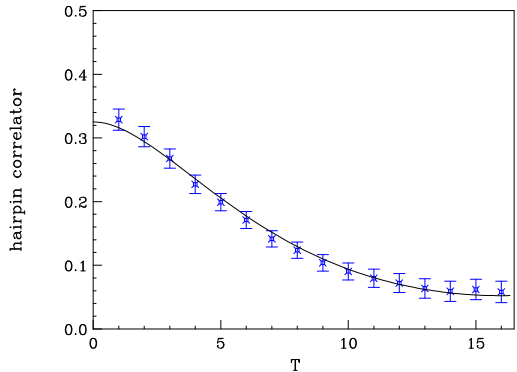}}
\vspace*{-.42in}
\caption{Hairpin propagator and double pole fit for $\kappa = .1394$.}
\label{fig:etap}
\end{figure}

In the quenched theory the disconnected part of the
eta prime correlator (hairpin term) has a double pole  
$ -m_0^2\over (p^2+m_{\pi}^2)^2$ 
whose coefficient determines $m_0$.
The correlator for $\kappa = .1394$ and a double pole 
fit($m_{\pi}= 0.261$ and $m_0 = 0.211$) is shown in Figure \ref{fig:etap}.
The $\delta$ parameter is given by:
\begin{equation}
  \delta = {m_0^2 \over 16\pi^2 f_{\pi}^2N_f}
\end{equation}

\begin{figure}
\epsfxsize = 0.42\textwidth
\centerline{\epsfbox{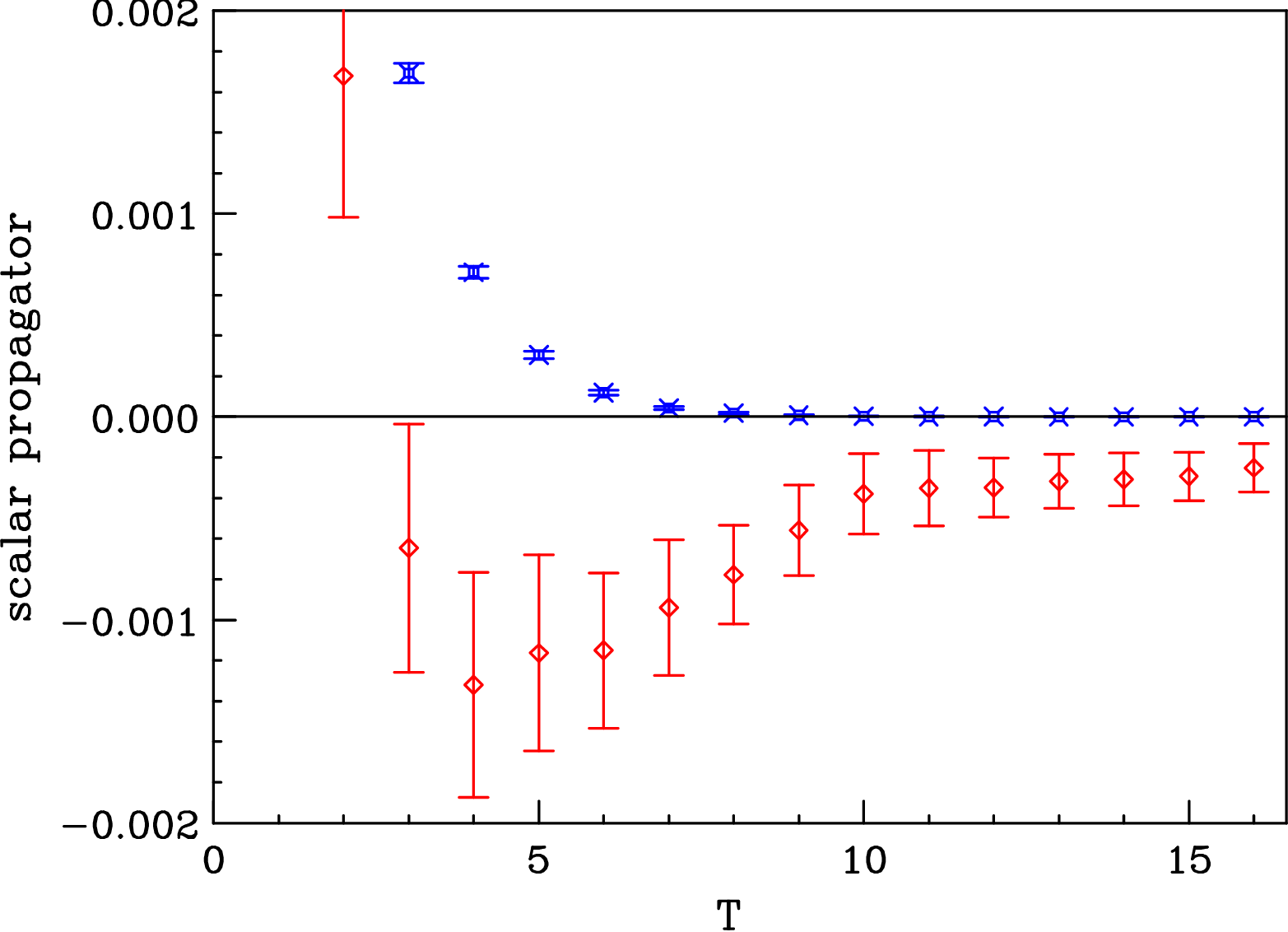}}
\vspace*{-.42in}
\caption{Isovector scalar correlators for $\kappa_{q} = \kappa_{\bar q} = .1397$
($\diamond$) and $\kappa_{q} = \kappa_{\bar q} = .1382$ ($\times$).}
\label{fig:scalar}
\end{figure}

As the quenched chiral limit is approached the
isovector scalar correlator shows negative norm behaviour.
The correlators for the lightest and heaviest $\kappa$ values are
shown in Figure \ref{fig:scalar}. 

Properly accounting for the effects of the 
hairpin - pion bubble\cite{scalar} allows a good fit of  
the isovector scalar correlator for all quark masses. 
One output is the ($a_0$) mass. 

\section{Results}

Using the MQA technique, meson properties (masses and decay constants)
can be extracted with sufficient accuracy to allow a fit of higher order
chiral parameters, $L_{5}$ and $L_{8}$. The physical results for 
$\beta = 5.9$ ($1/a|_{\rho} = 1.619$ GeV $Z_{A} = 0.865$) are compared 
with our $\beta = 5.7$ ($1/a|_{\rho} = 1.115$ GeV $Z_{A} = 0.845$) 
results\cite{chlogs}.
\begin{table}
\begin{centering}
\begin{tabular}{|c|c|c|}
\hline
 Parameter                &  $\beta = 5.9$     &  $\beta = 5.7$      \\
\hline
 $f=f_{\pi}$              &  $0.091(2)$  &  $0.100(2)$  \\ 
 $L_5 \times 10^3$        &  $1.55(27)$  &  $1.78(35)$  \\ 
 $(L_8-L_5/2)\times 10^3$ &  $0.02(5)$   &  $0.14(7)$   \\ 
 $r_0$                    &  $1.71(8)$   &  $1.99(12)$  \\ 
 $\delta$                 &  $0.053(13)$ &  $0.059(15)$ \\ 
\hline
\end{tabular}
\end{centering}
\end{table}

The physical values of $m_0$ (in GeV) extracted from the 
hairpin analysis (at $\beta = 5.9$) are $.526(14)$, $.533(14)$, $.554(14)$,
$.580(14)$, $.606(014)$ and $.613(17)$ for $m_{\pi} = .666, .612, .553, .492, 
.423$ and $.330$ GeV respectively. 
Extrapolating to $m_{\pi} = 0$ we obtain $m_0 = .642(15)$.
The corresponding value for $\beta = 5.7$ is  $m_0 = .548(25)$. 

Extracting $a_0$ and $a_1$ masses (at $\kappa_c$)
from scalar and axial vector propagators gives
$m_{a_0} = 1.33(5)$ and $1.34(6)$ GeV and 
$m_{a_1} = 1.27(4)$ and $1.12(8)$ Gev for $\beta = 5.9$
and $5.7$ respectively.
The value for $m_{a0} = 1.330(50)$ GeV suggests that the 
observed $a_0(980)$ resonance is a $K\bar{K}$ ``molecule'' and 
not an ordinary $q\bar{q}$ meson.



\end{document}